\documentclass[superscriptaddress,aps,prl,twocolumn,floatfix]{revtex4}
\usepackage{amsbsy,amssymb,amsmath,bm}
\usepackage{graphicx,color,ulem}
\newcommand{\f}[1]{Fig.~\ref{#1}}
\newcommand{\eq}[1]{Eq.~(\ref{#1})}

  \def\be{\begin{equation}}
\def\ee{\end{equation}} \def\bea{\begin{eqnarray}}
\def\eea{\end{eqnarray}} \def\l({\left(} \def\r){\right)}

\newcommand{\Tt}{T_{\text{th}}}
\newcommand{\Ht}{H_{\text{th}}}

\begin{document}

\title{Onset of dendritic flux avalanches in superconducting films}

\author{D.~V.~Denisov}
\affiliation{Department of Physics and Center for Advanced
Materials and Nanotechnology,    University of Oslo, P. O. Box
1048 Blindern, 0316 Oslo, Norway}
 \affiliation{A. F. Ioffe Physico-Technical Institute,
Polytekhnicheskaya 26,
St. Petersburg
194021, Russia}
\author{D.~V.~Shantsev}
\affiliation{Department of Physics and Center for Advanced
Materials and Nanotechnology,    University of Oslo, P. O. Box
1048 Blindern, 0316 Oslo, Norway}
 \affiliation{A. F. Ioffe Physico-Technical Institute,
Polytekhnicheskaya 26,
St. Petersburg
194021, Russia}
\author{Y.~M.~Galperin}
\affiliation{Department of Physics and Center for Advanced
Materials and Nanotechnology,    University of Oslo, P. O. Box
1048 Blindern, 0316 Oslo, Norway}
\affiliation{A. F. Ioffe
Physico-Technical Institute, Polytekhnicheskaya 26, St. Petersburg
194021, Russia}

\author{Eun-Mi Choi}
\affiliation{National Creative Research Initiative Center for
Superconductivity,
Department of Physics, Pohang University of Science and
Technology, Pohang 790-784, Korea}
\author{Hyun-Sook Lee}
\affiliation{National Creative Research Initiative Center for
Superconductivity,
Department of Physics, Pohang University of Science and
Technology, Pohang 790-784, Korea}
\author{Sung-Ik Lee}
\affiliation{National Creative Research Initiative Center for
Superconductivity,
Department of Physics, Pohang University of Science and
Technology, Pohang 790-784, Korea}

\author{A.~V.~Bobyl}
\affiliation{Department of Physics and Center for Advanced
Materials and Nanotechnology,    University of Oslo, P. O. Box
1048 Blindern, 0316 Oslo, Norway} \affiliation{A. F. Ioffe
Physico-Technical Institute, Polytekhnicheskaya 26, St. Petersburg
194021, Russia}

\author{P.~E.~Goa}
\affiliation{Department of Physics and Center for Advanced
Materials and Nanotechnology,    University of Oslo, P. O. Box
1048 Blindern, 0316 Oslo, Norway}
\author{A. A. F. Olsen}
\affiliation{Department of Physics and Center
for Advanced Materials and Nanotechnology,   University of Oslo,
P. O. Box 1048 Blindern, 0316 Oslo, Norway}
\author{T.~H.~Johansen}
\affiliation{Department of Physics and Center
for Advanced Materials and Nanotechnology,   University of Oslo,
P. O. Box 1048 Blindern, 0316 Oslo, Norway}
\email{t.h.johansen@fys.uio.no}

%\date{\today}

\begin{abstract}
%~\\[2mm]
We report a detailed comparison of experimental data and
theoretical predictions for the dendritic flux instability,
believed to be a generic behavior of type-II superconducting
films. It is shown that a thermo-magnetic model published very
recently~[\prb \textbf{73}, 014512 (2006)] gives an excellent
quantitative description of key features like the instability onset
(first dendrite appearance) magnetic field, and how the onset
field depends on both temperature and sample size. The
measurements were made using magneto-optical imaging on a series
of different strip-shaped samples of MgB$_2$. Excellent agreement
is also obtained by reanalyzing data previously published for Nb.
\end{abstract}


\maketitle

%Magnetic field penetrates type-II superconductors as
% Abrikosov vortices, which due to pinning build up a
%non-uniform metastable flux density distribution. Imaging methods
%have revealed that this state
% can break down through abrupt avalanches of flux forming
%complex dendritic patterns.

% INGRESS

% INTRODUCTION

Phenomena that create intriguing traces of activity that can be
observed by direct visual methods are among the most fascinating
things in nature. Penetration of magnetic flux in type-II
superconductors seen by magneto-optical (MO) imaging is one
example, where especially the spectacular dendritic flux patterns
occurring in superconducting films are currently attracting much
attention. The phenomenon has been observed in a large number of
materials; YBa$_2$Cu$_3$O$_x$, Nb, MgB$_2$, Nb$_3$Sn, NbN,
YNi$_2$B$_2$C and
Pb~\cite{Leiderer93,Duran95,Johansen01,Rudnev03,Rudnev05,jooss,menghini},
all films, and showing essentially the same characteristic
behavior. In abrupt bursts the film becomes invaded by flux in
narrow finger-like regions that often form a complex and
sample-spanning dendritic structure. These sudden events occur
typically during a slow ramping of the applied magnetic field, and
at temperatures below a certain fraction of the superconducting
transition temperature, $T_c$. It is also characteristic that the
flux patterns are never reproduced when experiments are repeated,
see \f{3colorMOI}, thus ruling out possible explanations based on
material defects guiding the flux motion. The massive experimental
data existing
today~\cite{Leiderer93,Duran95,Johansen01,Rudnev03,Rudnev05,jooss,menghini,Johansen02,Bolz03,welling04,Bobyl02,Barkov03,ye,Albrecht,Korea}
indeed suggest that the phenomenon is a generic instability of the
vortex matter in superconducting films.
\begin{figure}%[ttt]
\centerline{\includegraphics[width=8.5cm]{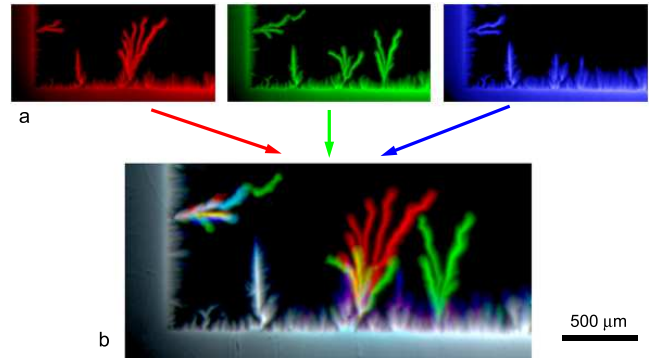}}
\caption{ {(a)}, Three MO images of flux penetration in
MgB$_2$ taken during repeated identical experiments. {(b)},
Image obtained by adding the 3 complementary colored images above.
In the sum image the grey tone regions are those of repeated
behavior, whereas colors show where there is no or only partial
overlap. Strong irreproducibility is seen in the dendrite shapes,
while the
 penetration near the edge and along static defects is
reproducible. The dendrites tend to nucleate at preferred sites
along the edge, which is due to small edge cavities giving local
field amplification. The experiments were performed after cooling
to 9.2 K and applying a magnetic field of 20 mT.
\label{3colorMOI}}
\end{figure}

Abrupt flux avalanches are known to occur in superconductors for
two fundamental reasons: (i) the motion of vortices releases
energy, and hence increases the local temperature, and (ii) the
temperature rise reduces flux pinning, and facilitates further
vortex motion. This makes up a positive feedback loop that may
lead to an instability~\cite{Mints81, Wipf91}. However, why such
avalanches should develop into dendritic patterns is a topic under
vivid discussion, and several competing theories were recently
proposed. They include a stability analysis taking into account
the complicating non-local electrodynamics of thin film
superconductors~\cite{Aranson,bigdoc}, a boundary layer model
assuming shape-preserving fronts~\cite{Baggio}, and a shock wave
approach~\cite{shapiro}, all leading to substantially  different
predictions. In this work we report on the first experiments
designed specifically to check the validity of these models. It is
shown that the model in Ref.~\onlinecite{bigdoc} provides an
excellent quantitative description of key features, such as the
instability threshold field, $\Ht$, i.e., the magnetic field when
the first avalanche occurs, and how $\Ht$ depends on both
temperature and the sample size. The results were obtained by MO
imaging of flux penetration in MgB$_2$ films. Also earlier
observations of Nb films~\cite{welling04} are shown to be in full
agreement with this model.

% EXPERIMENT

Thin films of MgB$_2$ were fabricated by a two step
process~\cite{Kang01}, where first a film of amorphous boron was
deposited on an Al$_2$O$_3$ $(1 \bar{1}  0  2)$ substrate
using a pulsed laser. The B film and high-purity Mg were then put
into a Nb tube, which was sealed in a high purity Ar atmosphere
and post-annealed at 900$^{\circ}$C. To eliminate possible
contamination with oxygen, water, and carbon, the samples were not
exposed to air until the final form of the film was produced. The
MgB$_2$ films possess $c$-axis orientation, as confirmed by
scanning electron microscopy, and magnetization data show a sharp
superconducting transition at 39~K. The film thickness was 300 nm.

A set of eight MgB$_2$ film samples was shaped by
photo-lithography into 3 mm long rectangles having different
widths ranging from 0.2 mm to 1.6 mm. All the samples were made
from the same mother film, allowing simultaneous and comparative
space-resolved magnetic observation \cite{remark}. An additional 5
mm wide sample was made using the same preparation conditions. A
standard MO imaging setup
%~\cite{Johansen96}
with crossed polarizers and a ferrite garnet indicator was used to
visualize flux distributions.
%
% RESULTS
%
\begin{figure}%[hhhh]
\centerline{\includegraphics[width=6.5cm]{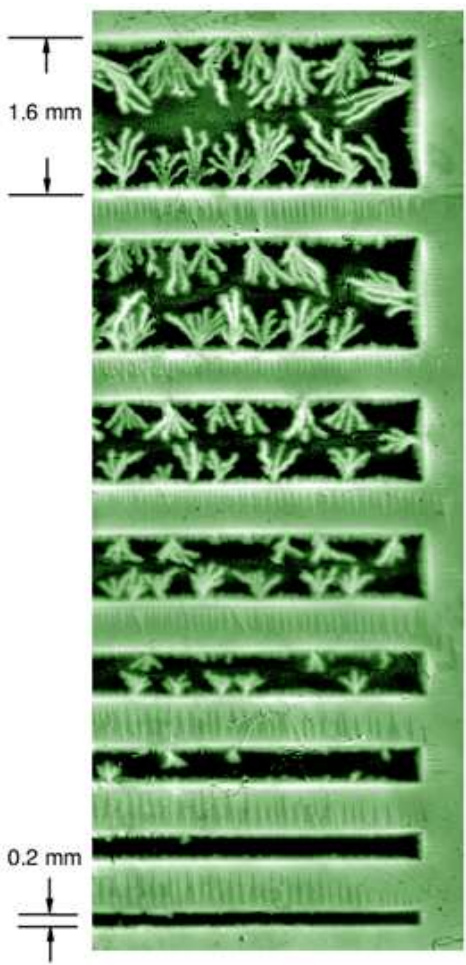}} \caption{
MO image showing flux distribution in MgB$_2$ strip-shaped samples at 4~K
and 15~mT applied field. The image brightness represents the local
flux density. Both the number and size of the dendrites are larger
for the wider samples. \label{MOImage}}
\end{figure}
Shown in \f{MOImage} is an image of the flux penetration pattern
when the eight samples, initially zero-field-cooled to 4~K, were
exposed to a perpendicular applied magnetic field slowly ramped to
15~mT. The magnetic flux enters the superconductor in a form very
much dominated by abrupt dendritic avalanches, although quite
differently for the various samples. It is evident that the number
of dendrites, their size and branching habit depend strongly on
the sample width. Whereas the wide strips become densely filled
with flux dendrites, the more narrow samples contain fewer, until
at the 0.2 mm wide strip flux dendrites almost never appear.

This qualitative result, was followed up by measuring how the
instability threshold field $\Ht$ depends on the strip width.
Results obtained for all eight strips are shown in \f{fig:H_w},
where each data point represents an average over 4 repeated
experiments using identical external conditions. The error bars
indicate the scatter in the observed  $\Ht$. A variation as much
as 30\% implies that the nucleation of this instability is
strongly affected by random processes, which is also consistent
with earlier
experiments~\cite{Duran95,Johansen01,Johansen02,welling04,Rudnev05,jooss}.
Nevertheless, the data in \f{fig:H_w} show a clear increase in the
threshold field as the strip becomes narrower. In other words,
reducing the sample width increases the stability of the
superconductor.

Measurements of the temperature dependence of $\Ht$ are shown in
\f{fig:H_Tfit}. One sees that $\Ht$ not only increases with
temperature, but appears to diverge at a certain temperature.
Above this threshold temperature, $\Tt$, found to be close to 10 K
for MgB$_2$ films, the dendritic instability disappears entirely.
Included in the figure are also data we have extracted from a
previous MO investigation of dendritic flux penetration in Nb
films~\cite{welling04}. The two behaviors show remarkable
similarities, although with different threshold temperatures,
approximately 6~K in the Nb case.
\begin{figure}[t]
\centerline{\includegraphics[width=8.5cm]{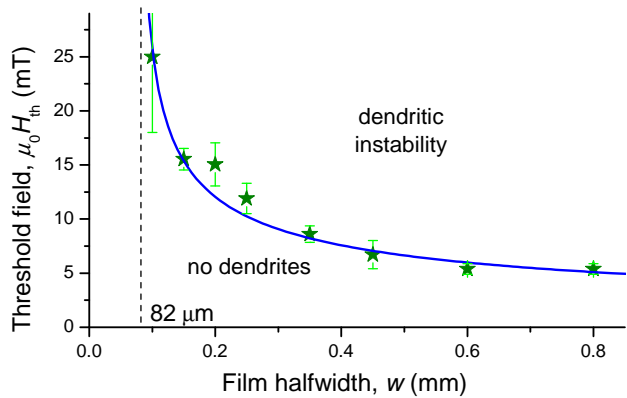}} \caption{
Threshold magnetic field for onset of the dendritic
instability in MgB$_2$ strips of different width
(symbols) plotted together with a fitted theoretical curve (full
line), which diverges at a finite $w$ indicated by the dashed
asymptote.\label{fig:H_w}}
\end{figure}

% DISCUSSION

To explain these observations we adopt the model developed in
Refs.~\onlinecite{bigdoc,Aranson}. There, a linear analysis of the
thermo-magnetic instability in a long and thin superconducting
strip thermally coupled to the substrate was worked out. The
analysis considered a strip of width $2w$ and thickness $d\ll w$
placed in an increasing transverse magnetic field leading to a
quasi-critical state in the flux penetrated region near the edges.
By solving the Maxwell and the thermal diffusion equations, it was
shown that for small fields there are no solutions for
perturbations growing in time, implying a stable situation. As the
field increases the distribution can become unstable, with a
fastest growing perturbation having a non-zero wave vector along
the film edge. This means an instability will develop in the form
of narrow fingers perpendicular to the edge -- a scenario closely
resembling the observed dendritic flux behavior. Within this
model, the threshold flux penetration depth, $\ell^*$, when the
superconducting strip first becomes unstable, is given by Eq.~(25)
of ref.~\onlinecite{bigdoc}, which can be expressed as
\begin{equation}
% \ell^* =  \frac{\pi}{2}\, \frac{\sqrt{n}\ L_d}{1-L_d/L_h}\, .
% % \quad d_y  =  \sqrt{ \frac{L_d L_h}{1 - L_d/L_h}}\, .
\ell^* =\frac{\pi}{2} \sqrt{\frac{\kappa }{|j_c^\prime| E}}
\left( 1 - \sqrt{\frac{2h_0 }{nd|j_c^\prime|E}}\right)^{-1}\, .
\label{k_xy}
\end{equation}
Here $j_c^\prime$ is the temperature derivative
of the critical current density,
% $T^*\equiv -(\partial \ln j_c/\partial T)^{-1}$,
%$L_h \equiv \sqrt{d\kappa/h_0}$ is the thermal healing length,
 $\kappa$ is the thermal conductivity, and
$h_0$ is the coefficient of heat transfer from the superconducting
film to the substrate.
%, i. e., the boundary condition at the interface
%is $\kappa \nabla T = h_0 (T-T_0)$, where $T_0$ is the substrate
%temperature.
% $L_d \equiv \sqrt{\kappa T^*/n j_c E}$ is a new length appearing
% the problem,
% where
 The parameter $n \gg 1$ characterizes the strongly nonlinear
current-voltage curve of the superconductor, described by the
commonly used relation for the electrical field, $E \propto j^n$.

\begin{figure}%[h]
\centerline{\includegraphics[width=8.5cm]{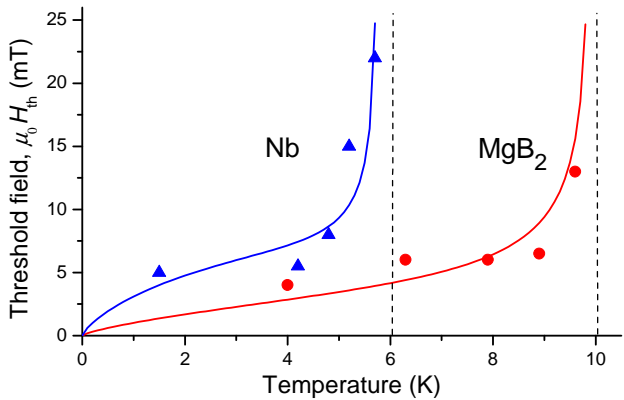}} \caption{
{Temperature dependence of the threshold magnetic field}.
Experimental data obtained for our largest MgB$_2$ sample and for
a 1.8 mm wide Nb film~\cite{welling04} are plotted as $\bullet$
and $\blacktriangle$, respectively. The full lines are theoretical
fits. The dashed lines show the limiting temperature above which
the instability vanishes. \label{fig:H_Tfit}}
\end{figure}
The threshold field, $\Ht$, is obtained by combining
Eq.~(\ref{k_xy}) with the Bean model expression for the flux
penetration depth of a long thin strip in a perpendicular applied
field~\cite{BrIn,zeld},
\begin{equation}
%\frac{\ell}{w}= 1- \frac{1}{\cosh \left({\pi H}/{ j_c d}\right)}
\Ht =  \frac{j_c d}{\pi} \;  {\rm arccosh} \left(\frac{w}{w-
\ell^*}
\right) .
\label{l_H}
\end{equation}
Plotted in Fig.~\ref{fig:H_w} as a solid line is this function
using $j_{c}=9\times 10^{10}$~A/m$^2$, a value obtained for
MgB$_2$ at 4~K by extra\-polation of $j_c(T)$-curves measured
under the stable conditions above $\Tt$. The only adjustable
parameter, $\ell^*$, was chosen equal to $82\ \mu$m, which gives
an excellent agreement with our data. It follows from \eq{l_H}
that narrower strips need a larger field to reach the critical
penetration depth $\ell^*$, which is exactly what we find
experimentally. Furthermore, the model predicts that $\Ht$ should
diverge when the strip halfwidth decreases towards $w = \ell^*$,
also this fully consistent with our MO observations.

To fit the observed $\Ht(T)$ one needs temperature dependent model
parameters. We assume then a cubic dependence of the thermal
conductivity, $\kappa=\tilde{\kappa} \l(T/T_c\r)^3$, as suggested
by low-temperature data for MgB$_2$~\cite{Gladun01}. Similarly, a
cubic dependence of the heat transfer coefficient, $h_0 =
\tilde{h}_{0} \l(T/T_c\r)^3$ is chosen in accordance with the
acoustic mismatch model confirmed experimentally for many
solid-solid interfaces~\cite{SwartzPohl}. Furthermore, we assume a
linear temperature dependence for the critical current density,
$j_c=j_{c0}(1-T/T_c)$, and with a pinning potential, $U \propto
1-T/T_c$, the exponent $n\sim U/kT$ also becomes $T$-dependent,
$n=\tilde{n}\;(T_c/T-1)$.

Combining all these equations,
one obtains a theoretical $\Ht (T)$, and shown in \f{fig:H_Tfit}
are such curves fitted to both sets of experimental data. The
model clearly reproduces the two key features;  (i) the existence
of a threshold temperature $\Tt$ above which the instability is
absent, and (ii) a steep increase of the threshold field $\Ht$
when $T$ approaches $\Tt$.
For MgB$_2$ the fit was made with $j_{c0}=10^{11}$~A/m$^2$, and
 $\tilde{\kappa}=160$~W/Km~\cite{Gladun01}, and choosing $\tilde{n}=10$
 corresponding at $T=10$~K to the commonly used $n= 30$.
The remaining parameters are the electric field and the heat
transfer coefficient, where best fit was obtained with $E=30$~mV/m
and $\tilde{h}_{0}=17$~kW/Km$^2$. It should be emphasized that the
experimental data for both $\Ht(w)$ and $\Ht(T)$ were fitted using
the same parameter values, and in both cases giving excellent
quantitative agreement. Figure~\ref{fig:H_Tfit} also shows a
similar fit for the data obtained for Nb, using $T_c=9.2$~K,
$j_{c0}=10^{11}$~A/m$^2$, $w$ = 0.9~mm, $d=0.5\
\mu$m,~\cite{welling04} $\tilde{\kappa}=120$~W/Km,~\cite{Koechlin}
$\tilde{n}=40$, $E=200$~mV/m and $\tilde{h}_{0}=36$~kW/Km$^2$.
Again the model excellently describes the experimental behavior.
\begin{figure}%[h]
\centerline{\includegraphics[width=8cm]{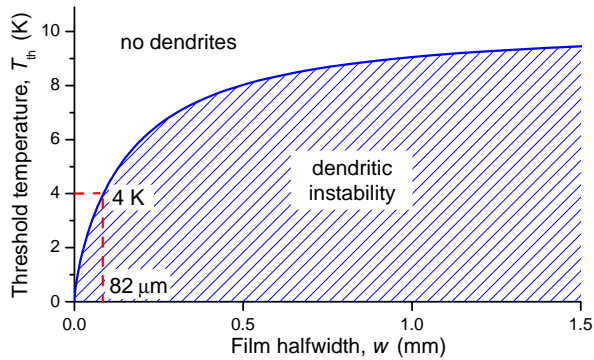}} \caption{
Theoretical stability diagram predicting the threshold
temperature $\Tt$ for different film width. The  curve is plotted
for parameter values corresponding to MgB$_2$ films.
\label{fig:T_w}}
\end{figure}

The fitted electric fields represent upper limiting values, since
we used bulk values for thermal conductivity, which in general are
larger than for films. Nevertheless, both  values largely exceed
the estimate, $E\sim \dot{H} \ell^*$, expected for a uniform and
gradual flux penetration with a ramp rate of $\dot{H} \approx $
1~mT/s as used in the experiments. We believe this discrepancy is
due to the fact that local, rather than global, conditions govern
the onset of the instability. Assuming that the flux dendrites are
nucleated by abrupt microscopic avalanches of
vortices~\cite{AltRev}, local short-lived electric fields can
easily reach those high values. In fact, such avalanches
consisting of $10^2-10^4$ vortices occurring in an area of $\sim
20\ \mu$m were recently observed by high-resolution MO imaging in
MgB$_2$ films~\cite{Shantsev05}. Electric fields close to 30~mV/m
would be created if such avalanches occur during a time span of
the order of $10^{-5}$ seconds. Randomness in such avalanches may
also explain the large scatter of the observed $\Ht$ values.
We also note that the estimated electric field at the {\it
nucleation} stage is still much lower than $E$ values at the tip
of an already {\it propagating} dendrite \cite{biehler}. This fact is in
agreement with expectations.

Finally, we emphasize that the two functions $\Ht (w)$ and $\Ht
(T)$ have a similar feature, namely a divergence at some value of
the argument beyond which the system becomes stable, see
Fig.~\ref{fig:H_w} and Fig.~\ref{fig:H_Tfit}. These stability
thresholds are actually related to each other by the condition
$\ell^*(\Tt)=w$. The relation between the threshold temperature
and the strip width is shown in~\f{fig:T_w}, and represents the
stability diagram in $w-T$ coordinates, here plotted for
parameters valid for MgB$_2$. It follows from the model that the
temperature range of the instability increases monotonously with
the strip width, but is limited upwards by a temperature close to
10 K for large-size films, as confirmed by many previous
experiments~\cite{Johansen01,Johansen02,Barkov03,ye,Albrecht}. The
general result that the instability is suppressed for sufficiently
narrow strips, is of particular importance for design of
superconducting electronic devices or other applications making
use of thin film superconductors operating at temperatures below
the instability threshold value.\\

% CONCLUSION

%In conclusion, we have presented a detailed comparison of
%experimental data and theoretical predictions for the dendritic
%flux instability in superconducting films. It was shown for the
%first time that a thermo-magnetic model can describe key features
%of the instability with an excellent quantitative agreement. This
%includes how the onset magnetic field for the instability depends
%on both temperature and sample size. In particular, it was shown
%how the instability is suppressed for sufficiently narrow strips,
%a result of general importance for design of superconducting
%electronic devices operating at temperatures below the instability
%threshold value.

%\begin{acknowledgments}
 This work is supported by the Norwegian
Research Council, Grant. No. 158518/431 (NANOMAT), and by
FUNMAT@UiO. We are thankful for helpful discussions with M.
Welling and D. Lazuko.
%\end{acknowledgments}


\begin{thebibliography}{99}
\bibitem{remark}
In principle, having several superconducting samples next to each
other leads to crosstalk, i.e., field expelled by one enhances the
 field experienced by another. This is a price to pay for
being able to make direct comparative observations. Our distance
between the strips was sufficiently large, as clearly seen from
\f{MOImage} where flux penetrates each strip equally from both
sides. Had crosstalk been important, substantial asymmetry would
be visible, especially in the upper sample having a neighbor only
on one side.
\end{thebibliography}
\end{document}